\documentclass[letter,12pt]{article}
\usepackage{graphicx,amssymb,amsmath,epstopdf,color,hyperref}
\usepackage{enumitem}

\def\beq{\begin{equation}}
\def\eeq{\end{equation}}
\def\bea{\begin{eqnarray}}
\def\eea{\end{eqnarray}}

\def\gev{\, {\rm GeV}}
\def\tev{\, {\rm TeV}}
\newcommand{\gsim}{\lower.7ex\hbox{$\;\stackrel{\textstyle>}{\sim}\;$}}
\newcommand{\lsim}{\lower.7ex\hbox{$\;\stackrel{\textstyle<}{\sim}\;$}}

\def\mpl{M_{\rm Pl}}

\def\msection#1{\section{\large #1}}

\definecolor{brightblue}{rgb}{0,0,1}
\begin{document}

\noindent
\begin{center}
{\bf\Large The Intrinsic and Extrinsic Hierarchy Problems} 

\vspace{0.3cm}
James D. Wells  \\ {\it Leinweber Institute for Theoretical Physics \\ University of Michigan, Ann Arbor} \\
\begin{small}jwells@umich.edu; ORCID 0000-0002-8943-5718\end{small}
%\today
\end{center}

\noindent
{\it Abstract:} 
The Hierarchy Problem of elementary particle physics can be divided into two separate problems: the Intrinsic and Extrinsic Hierarchy Problems. The Intrinsic Hierarchy Problem (IHP) arises when the Wilsonian renormalization group induces a large $\Lambda_{\rm UV}^2$ cutoff dependence on a much lighter scalar mass, creating a large finetuning. The Extrinsic Hierarchy Problem (EHP) arises when the IR theory is augmented with generically assumed extra states and interactions in the UV, making the resulting IR effective theory appear highly finetuned. The IHP is straightforward to analyze within a theory, but has suspicious regulator dependence, which has been suggested by some to be indication of a faux problem. The EHP is less straightforward to analyze, but has strength of physical intuition. We analyze EHP as a formal paradox, spelling out its premises and reasoning. From this we classify solutions to the EHP in terms of premise violations, and we articulate why some purported solutions to the Hierarchy Problem only partially solve the IHP and leave the EHP unaddressed.

\tableofcontents

\vfill\eject
%%%%%%%%%%%%%
\msection{Introduction}

The Standard Model of elementary particle physics has within its spectrum of particles the Higgs boson.  The Higgs boson is the only fundamental scalar boson in the theory, and as such its quantum self-energy corrections are quadratically sensitive to the cutoff  scale $\Lambda$ when viewed as an effective theory in the Wilsonian picture. This unique fact about the Higgs boson led to a realization that a pure scalar Higgs boson theory would suffer from a ``Hierarchy Problem": the quantum self-energy corrections with large $\Lambda$ would have to accomplish a very delicate finetuned cancellation with its bare mass in order to result in a small Higgs mass in the end, which is needed to described the elementary particle masses, all of which are below $175\gev$. The cutoff could be as high as the Planck scale $\Lambda\sim\mpl\simeq 10^{19}\gev$, which would generate a finetuning, based on this calculus, of one part in $10^{36}$. This constitutes a problem that many feel must be solved. It is expected, therefore, that some new physics would have to come in at a scale ($\Lambda_{\rm NP}$) near the weak scale ($M_{\rm weak}$) to limit the cutoff scale, $\Lambda\sim \Lambda_{\rm NP}\sim M_{\rm weak}$.

Solving the Hierarchy problem, and identifying the new physics that limits the cutoff scale in the Higgs boson corrections, has been a central motivation for many ideas of physics beyond the SM~\cite{Giudice:2008bi,Craig:2022eqo}. These include technicolor, composite Higgs, supersymmetry, and low-scale extra dimensional theories, to name a few categories. These new physics ideas are expected to manifest themselves near the weak scale in order to solve the problem. The hope was that the LHC would find evidence for one of these ideas, since its energy reach covered rather well the weak scale ($10^2\gev$), extending into the TeV scale ($10^3\gev$).

As we know, the LHC has found no evidence for new physics within its experimental reach. This might lead some to conclude that the Hierarchy Problem is a fake problem that has been debunked by empirical evidence. Nevertheless, one might reasonably hold that we face a conceptual crisis~\cite{Giudice:2017pzm}. However, the claim of this paper is that the Hierarchy Problem discussed above is just the first of two related but distinct hierarchy problems: the {\it Intrinsic Hierarchy Problem} and the {\it Extrinsic Hierarchy Problem}. 

The {\it Intrinsic Hierarchy Problem} is based on a rigid physical interpretation of the Wilsonian renormalization group as higher momentum modes of the theory are integrated out in the functional formalism. A theory is defined to be an effective theory up to some scale $\Lambda$ where bare parameters are defined. As one integrates momentum modes from the UV down to the IR there are quadratic coefficients to scalar fields $\phi$ generated:  $g^2_0\Lambda^2\phi^2$, where $g_0$ are couplings of the scalar field to itself and to other particles. Thus, for $\phi^2$ to have a small mass $m_\phi^2\ll \Lambda^2$, the bare mass parameter $m_0^2$ must  tune itself very delicately against $g^2_0\Lambda^2$ to accomplish this.  More details and historical reference are given below.

The second formulation of the Hierarchy Problem, the {\it Extrinsic Hierarchy Problem}, concerns itself with the stability of a very light Higgs boson in the presence of a multitude of other states in nature that can couple to it and raise its mass much higher than the weak scale. The {\it Extrinsic Hierarchy Problem} is  ``extrinsic"  since by its definition the problem does not exist within the SM itself but only possibly by the effects of states external to the SM. The {\it Extrinsic Hierarchy Problem} can be well cast as a paradox, with its premises, reasoning and absurd conflicting conclusions, all of which we will describe below.

In the following sections I first describe the historical development of the community's formulation of the Hierarchy problem. We show that the original formulation was decisively the {\it Intrinsic Hierarchy Problem}, with its enhanced physical attribution given to the dynamics of the Wilsonian renormalization group. We discuss some arguments that the {\it Intrinsic Hierarchy Problem} may be a faux problem after all, as has been suggested by some (see sec.~\ref{sec:extrinsic}), and that the second hierarchy problem, the {\it Extrinsic Hierarchy Problem}, is perhaps a more physical and immediate issue for particle physics. We formulate the {\it Extrinsic Hierarchy Problem} as a paradox to be resolved, and consider how various theories are organized as attacks on each of the premises of the theory. We also discuss how some theories that purport to solve the hierarchy problem are of only limited value in solving the {\it Intrinsic Hierarchy Problem} and completely ineffectual for solving the {\it Extrinsic Hierarchy Problem}. Conclusions from the analysis are given in the final section.

%%%%%%%%%%%%%%%%%%%%%%%%%%%%%%%%%%%%%%%%%%%%%%%%%%%%%%%
%%%%%%%%%%%%%%%%%%%%%%%%%%%%%%%%%%%%%%%%%%%%%%%%%%%%%%%
%%%%%%%%%%%%%%%%%%%%%%%%%%%%%%%%%%%%%%%%%%%%%%%%%%%%%%%
\msection{Two Formulations of the Hierarchy Problem}

\subsection{The Intrinsic Hierarchy Problem}

The Hierarchy Problem was first introduced by Susskind~\cite{Susskind:1978ms} upon reflection of the implications of the Wilsonian renormalization group applied to the Weinberg-Salam theory (i.e., the Standard Model's electroweak theory). Susskind's paper gets straight to the point:
\begin{quote}
{\it ``The need for fundamental scalar fields in the theory of the weak and electromagnetic forces is a serious flaw. Aside from the subjective esthetic argument, there exists a real difficulty connected with the quadratic mass divergences which always accompany scalar fields. These divergences violate a concept of naturalness which requires the observable properties of a theory to be stable against minute variations of the fundamental parameters."}~\cite{Susskind:1978ms}
\end{quote}
Susskind credits in the acknowledgements of this paper private communications with Ken Wilson for ``explaining the reasons why scalar fields require unnatural adjustments of bare constants."

The argument of Susskind, using slightly updated language, was that the renormalized Higgs mass parameter $m^2$, which should be of the weak scale ${\cal O}(10^2\gev)$, is related to its bare mass parameter $m_0^2$ and quantum corrections of order $g_0^2\Lambda^2$:
\beq
m^2=m_0^2+g_0^2\Lambda^2.
\label{eq:susskind}
\eeq
$\Lambda$ is the cutoff of the valid quantum field theory, and $g_0$ are couplings within the theory that arise in the quantum loop diagram computation, such as $g^2/16\pi^2$, $y_t^2/16\pi^2$, etc.
If the Weinberg-Salam theory is valid up to Planck scale of $10^{19}\gev$, then the cutoff of the theory is $\Lambda^2\sim M_{\rm Pl}^2$. In order for RHS of eq.~\ref{eq:susskind} to produce the LHS of $m^2\sim (10^2\gev)^2$, the bare mass parameter $m_0^2$ must be of order $M_{\rm Pl}^2$ to counter $g_0^2\Lambda^2$, and it must be tuned to one part in $10^{36}$ for the cancellation to be accomplished. As stated above, we call this formulation of the Hierarchy Problem the {\it Intrinsic Hierarchy Problem}, emphasizing that it is postulated to be a fatal flaw of any effective theory with fundamental scalar bosons existing significantly below the effective theory's validity cutoff scale.

Susskind did not argue precisely why such a cancellation is problematic, but only that ``such adjustments are unnatural and will be assumed absent in the correct theory." In a follow up paper with Dimopoulos~\cite{Dimopoulos:1979es} they reiterate this same discussion and gave a clearly stated characterization of the hierarchy problem:
\begin{quote}
{\it ``Indeed, it seems that to establish a hierarchy of mass scales, beginning at the Planck mass ($10^{18}\gev$) and ending at ordinary particle masses requires fundamental unrenormalized masses to be adjusted to 30 decimal places! Perhaps in some future theory such adjustments will appear natural, but at present divine intervention is the only available explanation."}~\cite{Dimopoulos:1979es}
\end{quote}
Dimopoulos and Susskind proceeded to describe theories of electroweak symmetry breaking that did not rely on a fundamental scalar boson, but rather condensates of fermions.

The dominance of this very physical interpretation of the Wilsonian renormalization group flow interpretation was further solidified in the influential quantum field textbook of Peskin and Schroeder~\cite{Peskin:1996,Peskin:2025lsg}, who gave a beautiful pedagogical description of how the Wilson renormalization group induces quadratic divergences to scalar masses proportional to the effective theory cutoff scale.  At the end of this discussion they echo the common interpretation of Susskind and many others over the years:
\begin{quote}
{\it ``Since the mass term, $m^2\phi^2$, is a relevant operator, its coefficient diverges rapidly under renormalization group flow. We have seen above that in order to end up at the desired value of $m^2$ at low momentum, we must imagine that the value of $m^2$ in the original Lagrangian has been adjusted very delicately.... [which] seems quite artificial when applied to quantum field theory of elementary particles, which purports to be a fundamental theory of Nature. This problem appears only for scalar fields.... Perhaps this is the reason why there seem to be no elementary scalar fields in Nature\footnote{This book was published in 1996 before the LHC found evidence for the scalar Higgs boson.}."}~(p.406 of~\cite{Peskin:1996})
\end{quote}
Thousands of quantum field theory students internalized this formulation of the {\it Intrinsic Hierarchy Problem}. Excellent analyses have been made on the Standard Model's potential intrinsic hierarchy problem within this framework from the perspective above and other perspectives~\cite{Polchinski:1983gv,deAlwis:2017ysy}. It is fair to say that the intrinsic hierarchy problem remains a key and legitimate concern within the community~\cite{Yamada:2020bqe,Branchina:2022jqc,Branchina:2022gll,Garces:2025rgn}.

In addition to key textbook discussions, Susskind and Dimopoulos's initial works were highly read and cited, and formed a basis of decades of further work on building theories of the Standard Model that solved the ``quadratic divergence" problem  in one way or another. Over the years they and others introduced technicolor, supersymmetry~\cite{Martin:1997ns}, composite Higgs~\cite{Panico:2015jxa}, Little Higgs~\cite{Schmaltz:2005ky}, Twin Higgs~\cite{Chacko:2005pe}, large extra dimensions~\cite{Arkani-Hamed:1998jmv}, warped extra dimensions~\cite{Randall:1999ee}, etc. All of these theories were constructed in the effort to solve the {\it Intrinsic Hierarchy Problem}, which naively looked catastrophic to the fundamental scalar theory of Weinberg-Salam unless radical new physics was invoked near the weak scale.

\subsection{The Extrinsic Hierarchy Problem}
\label{sec:extrinsic}

Cracks did develop in the community's unified view that the {\it Intrinsic Hierarchy Problem} was indeed a real problem that had to be solved\footnote{We could merely make an appeal to authority here by noting that the prophet of Wilsonian Renormalization Group, Dr.\ Wilson himself~\cite{Wilson:1983xri}, disavowed in 2004~(see sec.\,5 of \cite{Wilson:2004de}) his earlier claims that scalar bosons are unstable to high effective theory cutoffs. One must recognize, however, that his argument (some numbers are just ``unexpectedly small"!) did not address the central physical concern -- is the $\Lambda_{\rm cutoff}^2$ scale something to take seriously or not.}.
Several noticed that if one regulates the infinities of quantum field theory using dimensional regularization, there is not an obvious  finetuned cancellation anywhere in the Standard Model (see, e.g., \cite{Veltman:1980mj} and~\cite{Bardeen:1995kv} to name just two.) The cutoff $\Lambda$ is replaced with a dimensionless parameter $\varepsilon$ that regulates how many spacetime dimensions $D$ we are working in: $D=4-\varepsilon$. Infinities in the Higgs boson self-energy then show up as, for example, $\sim M_W^2/16\pi^2\varepsilon$, which do not evoke alarming cancellations. While this may not assure all that there is no  Intrinsic Hierarchy Problem~\cite{Branchina:2022jqc}, it does raise suspicions that it is a firm problem that must be solved.

Further developments, anticipated well by many, including~\cite{Veltman:1980mj}, \cite{Bardeen:1995kv} and~\cite{Gildener:1976ai}, emphasized that although cancellations between bare parameters and cutoff scale might arguably not be of much concern, a severe cancellation between parameters of beyond the SM theories, such as grand unified theories, are much more physically worrisome. We call this the {\it Extrinsic Hierarchy Problem}, which emphasizes the role of states external to the SM creating a need for highly finetuned cancellations in the formation of the low-energy SM effective theory. One way to express such worries in a somewhat unified way is the demand that there not be finetuned cancellations between high-energy theory parameters when matching to much smaller valued low-energy theory parameters  across heavy physical mass thresholds in effective theory matching conditions~\cite{Wells:2021zdp}. 

And so, historically, one has evolved from Susskind's claims about unacceptable cancellations between bare parameters and cutoff regulators (the {\it Intrinsic Hierarchy Problem}) to the additional claim that if there are heavy new states coupled to the Higgs boson the theory must not admit high finetuning across effective theory matching thresholds. The latter immediately becomes a problem (the {\it Extrinsic Hierarchy Problem}) when there exist many beyond the Standard Model mass thresholds to consider at high mass scales that couple to the Higgs boson. 

In contrast to the {\it Extrinsic Hierarchy Problem}, the original {\it Intrinsic Hierarchy Problem} claims appeared general and lethal under all circumstances for any theory that had a fundamental scalar with any particles coupled to it. The implication was obvious: fundamental scalars cannot exist or there needed to be radical new physics that guarantees internal cancellations of quadratic divergences. 
The more recent rise of the {\it Extrinsic Hierarchy Problem}, that new particles and interactions beyond the SM should not introduce extreme finetunings,  do not automatically de-legitimize fundamental independent scalars. The Standard Model itself simply has no Hierarchy Problem in this extrinsic formulation\footnote{This statement ignores quantum gravity considerations~\cite{tHooft:1979rat,Wells:2016luz}. For one, the cosmological constant has no known Natural explanation. Furthermore, quantum gravity may make the scale of quantum gravity ``physical", creating an {\it Extrinsic Hierarchy Problem} for the SM that requires solving.}. This is a relief since a legitimate theory of the Higgs boson, as seen and verified by experiment, cannot be ruled out by an controversial philosophical argument. However, this more recent {\it Extrinsic Hierarchy Problem} concern does  require care when considering beyond-the-SM theories so as not to introduce extreme finetunings. Any BSM theory that does introduce high finetunings across physical mass thresholds can reasonably be viewed as improbable~\cite{Wells:2018yyb}. It should be noted that the presence of an intrinsic hierarchy concern in a theory, signals that the theory is highly susceptible to an extrinsic hierarchy problem if there are mass thresholds at higher energy scales which activate the quadratic ``divergence" sensitivity. This is one way in which the two problems can be considered connected.

%%%%%%%%%%%%%%%
%%%%%%%%%%%%%%%%%%%%%%%%%%%%%%%
%%%%%%%%%%%%%%%%%%%%%%%%%%%%%%%
\subsection{Framing the Extrinsic Hierarchy Problem as a Paradox}

Our goal in this section is to describe precisely a formulation of the {\it Extrinsic Hierarchy Problem} in the logical presentation of a paradox, where the premises and reasoning are made explicit. 
As is often the case, making vague notions explicit runs the risk of generating disagreement on the details. There will be some who disagree on the precise formulation of the premises and the precise formulation of the reasoning, but that is precisely the purpose of this discussion. To make explicit what before was discussed nebulously is intended to clarify our thinking about possible resolutions and perhaps even other more trenchant reformulations of the hierarchy problem. 

In symbolic logic a paradox is represented by a set of beliefs or reasonably held premises $P_i$ that through acceptable rules of reasoning $R_\alpha$  leads to an unacceptable or absurd conclusion
\beq
\{P_i\},\{ R_\alpha\}\,\longrightarrow~Q,~~\text{where~}Q~\text{is~unacceptable}.
\eeq
The grounds for determining that $Q$ is unacceptable could be any of a number of possibilities. It could be religious faith, alternative lines of premises and reasoning that lead to $\neg Q$ (not $Q$), or, most often, $Q$ is in direct conflict with simple observation. 

 In the case of Olbers paradox, the premises (infinite universe, uniform density of stars, etc.)\ led to the absurd conclusion that the night sky is very bright. The conclusion is absurd because we observe that the night sky is dark. Zeno's paradoxes are often of the same form: seemingly unassailable premises and argumentation leading to absurd conclusions in conflict with observation, such as the conclusion in Zeno's Arrow paradox that motion is impossible.
 
Let us follow this approach and cast the issues of the {\it External Hierarchy Problem} into the structure of a Paradox. The sensible premises of the paradox will first be presented, and then a demonstration of the reasoning that leads to an absurd conclusion ($m_H\sim M_{\rm Pl}$) which is contradicted by observation ($m_H\ll M_{\rm Pl}$). What makes a paradox an especially rich one is when each of the premises must be true to yield the absurd conclusion, but one does not know which of the premises are false. This was true of Olbers paradox, and is true of the Hierarchy Paradox.

There are several ways one might formulate the Paradox's premises but the one that perhaps most conforms with the communities wrestling with the issue have these three premises:
\begin{enumerate}[label=P\arabic*]

\item {\bf Conventional Ur-Theory.} Nature is well described by an Ur-Theory just below the Planck scale with its Ur-Action and Ur-Langrangian density that is comprised of non-zero coefficients for all symmetry-allowed operators. The Ur-Theory is $3+1$ dimensional with standard spacetime symmetries already recognized (diffeomorphism invariance, Poincar\'e symmetry,) endowed with internal gauge symmetries characterized by $SU(3)\times SU(2)\times U(1)$ of the SM as well as other possible gauge symmetries and global symmetries applicable to the Ur-Theory's collection of conventional quantum fields and their interactions.

\item {\bf Aleatory Parameters.} Coefficients of the operators of the Ur-Theory are aleatorily assigned to each of the symmetry-allowed operators. The assigned parameters are given to the theory in the ultraviolet and have no teleological designs on the properties or implications of the theory in the deep infrared. 

\item {\bf Multitude of States.} Nature has many more scalars, fermions (both chiral and vectorlike), and vector bosons in its spectrum than just those of the Standard Model. In addition, many of these states have masses significantly higher than the weak scale, including masses near the highest mass scale of the Ur-Theory, just below the Planck mass.

\end{enumerate}

The Hierarchy Paradox says that if each of these premises is true then the Higgs boson mass should experience a condition where its mass $m_H^2$ is related to the difference between two (or many more) very heavy mass thresholds $M_{X}^2$ and $M_{Y}^2$ that must cancel:
\beq
m_H^2=M_X^2-M_Y^2.
\eeq
If $M_X^2$ and $M_Y^2$ are aleatory parameters that are very large $M^2_X,M^2_Y\sim M_{\rm Pl}^2$ it is an extreme unexpected finetuned coincidence (``divine intervention!" as Dimopoulos and Susskind might say) that they should cancel and give $M_X^2-M_Y^2\ll M_{\rm Pl}^2$. Thus, the paradox is born that the premises and the reasoning have lead to a conclusion ($m_H\sim M_{\rm Pl}$) that is absurd because it is in conflict with observation ($m_H\simeq 10^2\gev\ll M_{\rm Pl}$). 

The reasoning in this paradox included recognizing the implications of the premises, which includes assuming there are many new states (premise 3) coupling to the Higgs boson in standard ways (premise 1) with uncorrelated aleatory couplings (premise 2). In addition, our reasoning included the recognition that no extreme finetunings are allowed, which is sometimes called Naturalness.

To be more precise about this Naturalness reasoning, we consider the passing through energy thresholds from the Ur-Theory down to the SM where many effective theories are formulated along the way. These EFTs require solving matching conditions that match the low-energy theory parameters to those of the high-energy theory. Often these matching conditions take a form analogous to $Z_{\rm low}=X_{\rm high}-Y_{\rm high}$, as we discussed above. It is unexpected that $|Z_{\rm low}|\ll |X_{\rm high}|,|Y_{\rm high}|$, just as it is unexpected that the difference in the number of stars of two similarly sized galaxies should be many of orders of magnitude smaller than the total number of stars in the galaxies. Such statements have meaning because there is at least a plausible correlation that can be made between high finetuning of this type ($|Z_{\rm low}|\ll |X_{\rm high}|$) and very low probability~\cite{Wells:2018yyb}.

Summarizing the reasoning in the paradox we have three entries:
\begin{enumerate}[label=R\arabic*]
\item {\bf Logical Reasoning.} Correct standard reasoning applied to mathematics, language, logic rules, quantum field theory, etc.
\item {\bf Premises Acceptance.} Acceptance and implementation of the implications of each premise.
\item {\bf Naturalness.} No extreme unprincipled coincidence in couplings or finetuning across energy thresholds when passing from the UV to IR scales.
\end{enumerate}
By unprincipled coincidence or finetuning we mean one where there is no argument revealing its necessity. For example, the equivalent electromagnetic coupling of the electron and the muon is not an extreme coincidence since electromagnetic gauge invariance necessitates it (a principled coincidence). To be concrete, we consider a high finetuning to be $10^4$ and an extreme finetuning to be $10^{6}$~\cite{Wells:2021zdp}.

Let us now return to the premises and give some more detail and discussion to the meaning of each of them.

{\bf Ur-Theory.} The prefix Ur- is borrowed from the Old High German, which means original, ancient, or primordial. For us the Ur-Theory is the first sensible QFT parametrically lower than the Planck scale. It is the original theory in four dimensional spacetime, from which all the effective theories spring as one scales down in energy. The SM has its origin in a tumbling down of EFTs from the Ur-Theory to itself. We assume that all the EFTs below the Ur-Theory are formed in the IR from standard QFT techniques of integrating out states across their mass thresholds.
Within the Ur-Theory it is assumed that the principle of totality is at work, in that all operators that are allowed by the symmetries are present. There are no accidental zeroes for coefficients of the symmetry-allowed operators in the theory, although there can be by chance some values that are small (see next premise). 

{\bf Aleatory parameters.} The coefficients of operators in the Ur-Theory are assumed to be aleatoric. What that means is that they are randomly selected according to some principles and setting that is not entirely known to us, but the parameters are nevertheless contingent. They might arise from Wotan throwing dice or from quantum mechanical fluctuations on the moduli space of an Ur-Ur-Theory upon the birth of our universe. Perhaps we are a baby universe spawned in an eternal inflation scenario. The premise is agnostic to the precise mechanism, but it insists that the parameters of the theory are selected somehow, and that it has an element of contingency. And that when the parameters are selected there is no teleological force that tunes them precisely to satisfy some desired outcome in the IR. There might be an anthropic rationale for why we find ourselves in the particular universe we are in -- the premise does not address that -- but the selection of the parameters was contingent and without original purpose.

{\bf Multitude of States.} It is inconceivable to many that the only fields in all of nature are those that we have discovered up to now. Furthermore, it is inconceivable to many that the only fields that nature has in its spectrum are those that interact directly with humans. A generalized Copernican principle that we are not particularly special in the universe would suggest that nature is filled with many more states in its spectrum than the ones we know about. One can go beyond this general sensibility and note that in many theories that purport to unify forces or unify theories (quantum mechanics and general relativity) there are necessarily a large number of other fields that fill out nature's roster~\cite{Dijkstra:2004cc}. The premise holds that this is true. That many more states exist beyond just those we have already confirmed through experiment.

Given the discussion above, let us make a precise statement of the {\it Extrinsic Hierarchy Problem}:
\begin{quote}
{\it The Extrinsic Hierarchy Problem is the unresolved paradox with credible premises $\{P_1,P_2,P_3\}$ and sound reasoning $\{R_1,R_2,R_3\}$ that leads to the absurd conclusion that the Higgs boson mass should be many orders of magnitude above its measured value.}
\end{quote}

We have now formulated the {\it Extrinsic Hierarchy Problem} with some clarity that is intended to match the community's evolving sensibilities of quantum field theory and effective theories. We have identified the three premises to the Paradox and detailed the reasoning features in the paradox. Both the premises $\{P_1,P_2,P_3\}$ and the reasoning $\{R_1,R_2,R_3\}$ are intended to be recognized by the community as a plausible articulation of the paradox that we face after the discovery of the Higgs boson. Like all paradoxes, resolutions occur by critical evaluation of both the premises and the reasoning, which is what we turn to next.

%%%%%%%%%%%%%%%%%%%%%%%%%%%%%%%%%%%%%%%%%%%%%%%%%%%%%%%
%%%%%%%%%%%%%%%%%%%%%%%%%%%%%%%%%%%%%%%%%%%%%%%%%%%%%%%
%%%%%%%%%%%%%%%%%%%%%%%%%%%%%%%%%%%%%%%%%%%%%%%%%%%%%%%
\msection{Resolutions to the Extrinsic Hierarchy Problem}

%%%%%%%%%%%%%%%%%%%%%%%%%%%%%%%%%
\subsection{Violating premise 1: conventional Ur-theory}

When trying to resolve the Hierarchy Problem a key approach has been to call into question the ``conventional Ur-theory" premise. It is in attacking this premise that supersymmetry's status rose to become by far the most popular approach.

\noindent
{\it Supersymmetry}

Supersymmetry violates the paradox by not accepting that the only symmetries possible in the Ur-theory are analogous to the symmetries we already know. Supersymmetry, instead, can be interpreted as adding extra fermionic-like (anti-commuting) dimensions to spacetime. Supersymmetry extends spacetime into superspace with coordinates $(x^\mu,\theta^\alpha,\bar\theta^{\dot\alpha})$, where $x^\mu$ are the ordinary $3+1$-dimensional ordinary (commuting) spacetime coordinates and $\theta^\alpha,\bar\theta^{\dot\alpha}$ are Grassman-valued spinorial (anti-commuting) coordinates, where $\theta^\alpha\theta^\beta=-\theta^\beta\theta^\alpha$.

Superfields, which are the core quantum fields of a supersymmetric theory, are functions of these superspace variables $\Phi(x,\theta,\bar\theta)$, analogous to ordinary fields being functions of just the spacetime variables $\phi(x)$, $\psi(x)$, $A^\mu(x)$, etc. Just as $P_\mu\sim -i\partial_\mu$ generates translations in $x^\mu$ in the Poincar\'e group, the supersymmetric generators $Q_\alpha,\bar Q_{\dot\alpha}$  are fermionic derivatives on the $\theta$ variables and generate translations in the fermionic directions $\theta$ of superspace. 

Annihilating the premise that the Ur-theory is conventional is not enough to solve the Hierarchy problem, but it was soon discovered that softly broken supersymmetric theories could do exactly that. Supersymmetry cannot be perfectly preserved in nature otherwise the electron (spin $1/2$) and its superpartner (selectron, spin $0$) would have the same mass and supersymmetry partners would have already been found. Instead, supersymmetry must be spontaneously broken to raise the superpartner masses to beyond experimental reach. In this case an ordinary particle and its superpartner have equal mass up to supersymmetry breaking mass. 

An example of how softly broken supersymmetry operates, we examine the special case of the top quark. The mass of the top quark $m_t$ and top squark (stop) superpartner  $\tilde m_t$ are
\beq
m_{\rm top}=m_t,~~~{\rm and}~~~m_{\rm stop}^2=m_t^2+\eta_t\tilde m^2
\eeq
where $m_t=173\gev$ (the known top quark mass), $\tilde m^2$ is the unknown supersymmetry breaking mass squared, and $\eta_t$ is an ${\cal O}(1)$ parameter that relates supersymmetry breaking mass $\tilde m$ to each field of the theory. 

One can then compute the self-energy correction of the Higgs boson mass due to the top and stop contributions, and one finds that quadratic divergences cancel and one is left with 
\beq
\Delta m_H^2\propto -\frac{y_t^2}{16\pi^2}\left[ (m_t^2)-(m_t^2+\eta_t\tilde m^2)\right]\propto \frac{y_t^2}{16\pi^2}\tilde m^2.
\eeq
Therefore, as long as $\tilde m\lsim {\rm TeV~scale}$ there is no large quantum contribution to the Higgs mass. Furthermore, one can add any particles at any scale and the magic of supersymmetry will always yield a similar result that the quantum corrections to the Higgs mass squared all cancel up to factors of $\tilde m^2$.

The reader may note that the above discussion appears on the surface to be primarily about solving the {\it Intrinsic Hierarchy problem} by canceling $\Lambda^2$ quadratic divergences within the theory. However, the relevance of supersymmetry to the {\it Extrinsic Hierarchy Problem}, which is our primary paradoxical concern here, is that any new states introduced will not allow any quadratic divergences to occur up to the supersymmetry breaking scale. It is in that sense that the paradox could have been in principle resolved.

However, there are experimental consequences to consider. The solution suggests that $\tilde m^2$ should be near the weak scale. Thus, there was much hope that the LHC would find evidence for supersymmetry. It did not. It is somewhat controversial to decide whether the LHC definitively has ruled out supersymmetry as a resolution to the Hierarchy Problem or if there is still some Natural, non-finetuned room for a supersymmetric theory to be the answer~\cite{vanBeekveld:2019tqp,Baer:2020kwz,Tata:2020afe}.

It is interesting to contemplate for a moment that from merely a low-energy theory perspective with only the immediacy of the theory right in front of us supersymmetry can only be a problem. It could introduce finetunings that the Standard Model just did not have. These finetunings could be from very high superpartner masses or the doublet-triplet splitting problem in a supersymmetric grand unified theory, or finetuning in the solution to the $\mu$-problem of supersymmetry.
Only upon considerations of more expansive premises about the nature of physics from the weak scale to the Planck scale do these legitimate worries arise, as captured by the {\it Extrinsic Hierarchy Problem}. Nevertheless, supersymmetry is or was an excellent contender for solving the paradox, depending on nature's tolerance level for finetuning.

\noindent
{\it Extra spatial dimensions}

One does not have to reach into fermionic extra dimensions to violate the conventional Ur-theory premise and resolve the {\it Extrinsic Hierarchy Problem}. One can merely introduce extra spatial dimensions on top of the three that we already experience. In order to be consistent with ordinary observations these extra spatial dimensions need to be compactified, meaning that their existence is only perceived as one goes to much shorter distances and higher energies.

One approach was to have very large extra dimensions compactified with a radius $R$ whose value is significantly larger than the fundamental length scale of the theory. For example, in the ADD model~\cite{Arkani-Hamed:1998jmv},  if there are $n$ extra dimensions of size $R$ then the relation between $R$, $M_{\rm Pl}$, and  the true highest fundamental scale of the theory $M_D$ is
\beq
M_{\rm Pl}^2\sim R^n M_D^{2+n}.
\eeq
In this scenario the quantum corrections to the Higgs boson mass are all bounded by the new fundamental scale $M_D$ which can be many orders of magnitude below the Planck scale. The Planck scale is no longer a physical scale, but rather an effective scale for the graviton interactions, which spread into the extra dimension, thereby lowering their strength compared to what a $3+1$-dimensional gravity theory with fundamental scale $M_D\sim M_{\rm weak}$ would give. 

The ADD theory is probably the simplest and most violent destruction of the conventional Ur-theory premise in that it eliminates the proximate problem of the large hierarchy altogether. It does come with a price, which is how did the extra spatial dimensions get such a large size. 

In a similar spirit one can introduce extra spatial dimension(s) that are warped in factorizable metric geometry~\cite{Randall:1999ee}. If $x^\mu$ are the ordinary $3+1$ dimensional spacetime coordinates, and $y$ is a fifth dimension, one could consider the consequences of a spacetime metric of the form
\beq
ds^2=e^{-2k|y|}\eta_{\mu\nu}dx^\mu dx^\nu+dy^2
\eeq
This forms an ${\rm AdS}_5$ space with curvature scale $k$. The exponential warp factor, $e^{-2k|y|}$, whose absolute value implies $S^1/Z_2$ orbifold symmetry (identify $y\to -y$), is the key to its relevance to the  Hierarchy Problem. 

If the Higgs boson lives on the $y=\pi r_c$ brane, then any mass scales on that brane are suppressed by $e^{-k\pi r_c}$, where $r_c$ is the orbifold compactification radius. One only needs a factor of $kr_c\sim 12$ or so to suppress a mass from $m_0$ down to $m_{\rm phys}\sim {\rm TeV}$. There are no finetunings in numbers like $\sim 12$ and given that all the physical masses on the SM brane are $\sim {\rm TeV}$ and cannot be higher, there are no quantum corrections to the Higgs boson mass touching $\sim M_{\rm Pl}^2$.

\noindent
{\it Conformal symmetry}

At the classical level the Standard Model is conformally invariant except for the mass term of the Higgs boson operator. Thus, an obvious hope would be to construct a conformally invariant theory at leading order which then breaks weakly to produce a small Higgs boson mass well below the Planck scale. However, it is notoriously difficult to produce a respectable working theory directly using this approach, despite serious attempts at doing so (see, e.g.,~\cite{Frampton:1999yb,MarquesTavares:2013szc}). Nevertheless, work and progress continues in this promising direction~\cite{deBoer:2024jne}.

It has also been recognized that through the AdS/CFT correspondence the Randall-Sundrum extra-dimensional approach has a dual description as a (near) conformal field theory~\cite{Arkani-Hamed:2000ijo}. There is still much to explore with respect to this equivalence, and much to explore regarding a conformal field theory origin to the small Higgs mass problem~\cite{Blasi:2022hgi}. In any event, a CFT, or related ideas~\cite{Foot:2013hna,Clarke:2015bea,Aoki:2012xs}, would be violating the conventional Ur-theory premise, given its special symmetry-protection character.

\noindent
{\it Lack of LHC discovery}

One must recognized that the lack of evidence of these alternatives to conventional Ur-Theories at the LHC has reduced interest in them over recent years. The hope was that the necessity of no large finetunings within these new theories in order to create the low-energy Standard Model would force some of their visible states to be in the discovery reach of the LHC. That has not happened, and it is interesting to ask why. 

The most obvious answer would be to say that they are just not what nature has chosen, and we must be clever and think of new ideas. Another approach is to suggest that the LHC was not terribly high energy anyway. The limits on the masses of the superpartners of supersymmetry are only about one order of magnitude heavier than the Higgs boson mass, and certainly not two orders of magnitude higher. Given such low mass reach, our expectation of supersymmetry showing up at the LHC was more a hope rather than a rigorous theorem.

 To rephrase the point made above, one could reasonable say that the hope for finding supersymmetry at the LHC was based on answering the subconscious question, ``What is the lowest superpartner mass that makes me begin to squirm about finetuned cancellations in the electroweak sector?" The answer to that question is about $1\tev$, within range of LHC experiments. However, a more relevant question to determine how dispositive the LHC has been to the viability of supersymmetry is, ``What's the highest supersymmetric partner mass where I am sure that Nature would find unacceptable finetunings?" Any reasonable approach to that question, perhaps PeV scale, would yield masses rather comfortably above the searching probes of the LHC. 
 
 Similar kinds of hope-versus-rigor considerations can be applied as well to extra dimensional theories, composite Higgs theories, etc. Unfortunately the LHC energies are not enough to rule out most of the new theories that have been discussed, but it is fair to say that having not shown up in prime real estate energy levels (below $\sim 1\tev$) has reduced the posteriors for these theories.

%%%%%%%%%%%%%%%%%%%%%%%%%%%%%%%%%
\subsection{Violating premise 2: Aleatory parameters}

The most difficult premise to articulate in a scientific manner is the aleatory premise, which states that the parameters of the Ur-Theory are contingent, being randomly selected over perhaps a (near) infinite number of possibilities. 

\noindent
{\it Wotan throwing dice}

One concrete formulation of this premise is that the parameters of the Ur-Theory in our particular universe were born by a random, quantum-mechanical selection on a landscape of an enormous number of  vacua~\cite{Douglas:2003um}, which is one of many universes spawned perhaps in eternal inflation~\cite{Vilenkin:1983xq,Freivogel:2005vv,Linde:2009ah}.

In other words, the premise assumes that somehow the specifics of our universe are contingent and not necessary, as though the Norse god Wotan threw dice to determine coupling constants, if not also gauge symmetries, number of spatial dimensions, etc. Wotan's dice throwing could be shorthand for the spawned baby universes discussed above, or some other as-yet unknown source of contingency.

\noindent
{\it Necessity of the premise}

Somehow this contingency premise feels intuitive to most scientists, although if asked to prove it, or to even take the premise to be provisionally absolute in order to investigate implications, it becomes very uncomfortable. Nevertheless, {\it without the aleatory premise the Hierarchy Problem becomes meaningless.} The only way that the Hierarchy Problem can be a problem is if it is viewed probabilistically, or absolutely, impossible to have $m_H\ll M_{\rm Pl}$. If there is no absolute principle forbidding $m_H\ll M_{\rm Pl}$, which there decisively is not, then it all becomes a probability question. Without a probability density existence supposition of one kind or another the exercise is meaningless. 

Realizing the necessity of the premise reveals an important possible resolution of the Hierarchy Problem: there is no Hierarchy Problem if there is no contingency. We are the one and only universe, not born in a distribution of many possibilities. The parameters were not determined in a dice game between Wotan and his friends. They are just so. And if they are just so, there can be no discussion of improbable. Declaring a special value to be improbable in such circumstances would be committing a fallacy of illicit probabilistic inference.

If one declares that all aspects of our universe are simply necessary and not contingent from some distribution the game is up. The Hierarchy Problem is no more. Likewise, the doublet-triplet splitting problem in grand unified theories should be of no concern. Nor should the cosmological constant problem. Nor the baryon asymmetry problem. Nor any extremely flat inflationary potential. None of these are meaningful concerns anymore if we take the viewpoint of absolute necessity of the universe with no contingency\footnote{One could try to turn the logic around and use those intuitions to meekly prove (``proof by useful intuition") that the universe is contingent.}.

However, our intuition appeals to contingency for some reason that we can only recognize but not rigorously dissect. This intuition ranges from how we think the universe was put together to how we weigh alternative defense theories in a criminal trial.  At this point we just acknowledge that the aleatory premise must be at work if the Hierarchy Problem remains a paradox to be resolved, which we do.

\noindent
{\it Teleological designs}

Part of the aleatory premise is that there are no preconceived, or teleological designs on the selection of parameters of the Ur-Theory. In other words, when Wotan is throwing dice to determine the fields and their couplings he does not manipulate the game with a purpose in mind for the final outcome in the deep infrared.

An example for  teleological designs to be effectively present, although not in principle present, is the anthropic principle. The anthropic principle says that whatever parameters and theories have developed they must take on values that allow the forming of intelligent life to discuss it (us). So although Wotan can throw dice all he wants, the universe he creates will have nobody to talk about it unless certain conditions are met.

The anthropic principle has a long and distinguished literature~\cite{Barrow:1988}. Its most prominent place of activation is in the discussion of the cosmological constant problem. The cosmological constant is many orders of magnitude below the naive quantum field theory value of at least $m_t^4$. After decades of work on the problem there are no good solutions. And so the resolution may very well be that the cosmological constant had to be as low as it is in order for structure to form~\cite{Weinberg:1987dv}, which is a precursor to development of intelligent life to discuss it. There are many technicalities and many caveats to the discussion~\cite{Adams:2019kby}, but the fact remains no other known approach to the cosmological constant problem is nearly as promising as the anthropic one. 

Might the light Higgs boson be solved through the anthropic principle also? Let us briefly discuss the similarities and differences of the cosmological constant problem and the hierarchy problem. The similarity is that the cosmological constant is a dimension-zero operator that gains $\Lambda^4$ contributions in the Wilsonian renormalization group flow from its cutoff of $\Lambda$ to the IR. These quartic divergences are analogous to the quadratic divergences in front of the gauge invariant and Lorentz invariant $|H|^2$ operator of the Higgs boson. 

There are two key differences though. The cosmological constant obtains corrections of order $M_{\rm SM}^4$ from SM particles, which are physical and alone are large enough to completely overwhelm the measured cosmological constant. This is in contrast to the Higgs boson mass squared picking up corrections of order loop factor times $M_{\rm SM}^2$ which {\it do not} overwhelm the Higgs boson mass and are Natural. Thus, the cosmological constant problem suffers immediately a pressing physical conundrum intrinsic to SM+gravity itself. However, unlike the SM Higgs problem, the cosmological constant problem is tied up with gravitation. Without gravity the zero point function value would have no relevance. Quantum gravity is a severe mystery.  It clearly has complex IR/UV mixing due to, for example, ever increasing higher energy (UV) collisions forming larger and larger (IR) black holes. One can be agnostic about the severity or not of a problem that is deeply tied to gravitational effects, and suggest, as 't Hooft did that it might be reasonable to suppose that ``{\it only} gravitational effects violate naturalness" (emphasis are 't Hooft's)~\cite{tHooft:1979rat}.

There remains another difference between the cosmological constant problem and the Higgs mass problem that is relevant to this discussion. It is now well understood how the anthropic principle might be connected to a very low cosmological constant. However, after years of trying, no obvious connection has been made between a light Higgs boson and the anthropic principle~\cite{Weinberg:2000yb}. Until a correlation or explanation can be found, one is best served looking elsewhere.

The best approach to tackle the teleological aspect of the aleatory premise is to investigate theories that indeed do have firm ultraviolet-infrared connections. This becomes a deeper theoretical exercise that is not about conventional model building, but rather upending the Wilsonian paradigm of quantum field theory. The Wilsonian approach to quantum field theory, with its insistence on dramatic independent separation of UV and IR degrees of freedom, may be what is holding us back. Most of us have swum in Wilsonian waters all our lives and we may need  to wiggle ourselves onto land to resolve our conceptual Hierarchy problems.

There have been some inchoate studies in various fundamental approaches to UV/IR connection. Some ideas are inspired by string theory, which necessarily has some UV/IR connection such as $T$-duality where short distance are connected to long distances via $R\to \alpha'/R$. Semi-explicit proposals have been made that purport to make headway in solving the hierarchy problem through this method (see, for example, \cite{Abel:2024twz}). These usually target the {\it Intrinsic Hierarchy Problem}. Others ideas are inspired by noncommutative field theory ($[x^\mu,x^\nu]\neq 0$) ~\cite{Chamseddine:2012sw,Craig:2019zbn}. And of course AdS/CFT correspondence is a prime inspiration since the UV physics of the boundary conformal field theory corresponds to the IR physics in the bulk AdS spacetime, and vice versa~\cite{Arkani-Hamed:2000ijo}. All of these ideas are quite theoretical and speculative, and constitute radical departures from ordinary field theory that has been successful for so long in describing nature. Nevertheless, if one is convinced that the Wilsonian renormalization group does give an {\it Intrinsic Hierarchy Problem} in addition to the {\it Extrinsic Hierarchy Problem}, such radical ideas may be the only path to solution.

%%%%%%%%%%%%%%%%%%%%%%%%%%%%%%%%%
%%%%%%%%%%%%%%%%%%%%%%%%%%%%%%%%%
%%%%%%%%%%%%%%%%%%%%%%%%%%%%%%%%%
\subsection{Violating premise 3: Multitude of States}

The third premise states that there is a multitude of additional states beyond the SM of all kinds (scalars, fermions, and vector bosons) at a multitude of scales. There is no proof that such a statement is true, but there is some circumstantial evidence for it.

\noindent
{\it New particles and interactions expected}

There are many puzzles in nature that the SM alone does not seem to be able to solve. For example, the laws of physics and our conception of the early universe's evolution does not appear to be compatible with the current dominance of matter over antimatter in the universe. If the particle content of the universe arose out of the production of states from an extreme source of energy, there should be equal numbers of baryons as antibaryons. Generating this asymmetry is called baryogenesis, and the widely accepted view is that new states and new interaction symmetries, etc.\ are needed for baryogenesis.

Approximately 27\% of the total energy density of the universe today appears to be made of dark matter. The required properties of dark matter are that it cannot interact strongly with baryons, and must form a more diffuse halo around galaxies than the clumping strongly interacting baryonic matter. With the possible and remote exception of primordial black holes~\cite{Carr:2020xqk,Green:2020jor}, the SM would need to be augmented with new states to account for this dark matter.

There are other data-driven reasons, such as inflation and neutrino mass generation, why the SM particles are unlikely to be all there is in nature. A somewhat more speculative reason why additional states are expected is grand unification, which posits that the three gauge couplings of the three gauge forces, electromagnetism, weak force and strong force, meet and unify at some high-energy scale, perhaps a few orders of magnitude below the Planck scale. A grand unified theory would surely add many more scalars and vector bosons to the spectrum. In the grand unified theory case, the urgency of the Hierarchy Problem takes its most concrete form, with most model building putting these states at $10^{14-16}\gev$, many of which interact directly with the Higgs boson. The famous doublet-triplet splitting problem of minimal $SU(5)$ is a tree-level instantiation of the {\it Extrinsic Hierarchy Problem}, which attracted great attention in the theory community in the early days of GUT model building.

Finally, a Copernican-like outlook on nature also suggests that we are not terribly special in the universe and that all the fundamental particles of nature are not just those that interact directly with us. A more expansive view would admit the strong possibility that there is more to the universe. 

We note in passing here that one could go to the other extreme and postulate the hierarchy problem being solved by an extreme multitude of extra states~\cite{Dvali:2009ne,Arkani-Hamed:2016rle}. In a sense this violates the multitude of states premise, because these solutions require a deluge of states well beyond standard conceptions of particle multiplicities in model building. 

\noindent
{\it No scalar Higgs boson}

From the considerations above, it appears that this premise is not likely to be violated. However, there is a sense in which it could be violated minimally that is directly related to resolving our Hierarchy paradox, and that is that electroweak symmetry breaking is not accomplished by a fundamental scalar boson.

The first attempt at solving the Hierarchy Problem was to posit that the electroweak symmetry breaking is accomplished by a condensate of exotic fermions. Nature demonstrates this type of dynamical symmetry breaking in many circumstances. Superconductivity is the condensate of electron Cooper pairs $\langle ee\rangle$. Chiral symmetry breaking in QCD is the condensate of light quarks $\langle q^\dagger_Lq_R\rangle$, which happens to also break the $SU(2)\times U(1)$ electroweak symmetry breaking since $q_L^\dagger q_R$ has the same quantum numbers as the Higgs boson, being a doublet under $SU(2)$ and hypercharge value of $1/2$. The vacuum expectation of this condensate is not large enough to account for the experimental known value needed, but that does not detract from the mechanism's validity and precedence.

If one adds exotic heavy fermions ($T_L$, $T_R$, etc.)\ to the spectrum a condensation scale much larger might be possible through $\langle T_L^\dagger T_R\rangle$. This approach is straightforward to implement to achieve the full $W^\pm$ and $Z^0$ boson masses, but there were persistent unresolved difficulties to account for the fermion masses of the SM through this approach without being in conflict with a host of other experimental data, such as flavor-changing neutral current measurements. 

In addition to the most straightforward Technicolor schemes for electroweak symmetry breaking there were more sophisticated approaches to model building that gave rise to ``Higgsless theories"~\cite{Csaki:2003zu} and composite Higgs boson theories~\cite{Panico:2015jxa} that did not rely on a fundamental scalar boson in the spectrum. 

Most of the proposed theories of non-fundamental scalar boson Higgs mechanism were ruled out rather quickly once the Higgs boson was discovered in 2012. Of course, there was always the possibility that the Higgs boson discovered at CERN is not a fundamental scalar but a composite scalar, analogous to a pion. However, the remarkable correlation of the Higgs boson decay rates into its various final states ($H\to \bar b b$, $\tau^+\tau-$, $\gamma\gamma$, $ZZ^*$, $WW^*$, etc.)\ with the predictions of the fundamental scalar Higgs boson theory of the SM~\cite{Carena:2023} are putting increasing finetuning pressures on non-fundamental Higgs boson theories.

%%%%%%%%%%%%%%%%%%%%%%%%%%%%%%%%%
%%%%%%%%%%%%%%%%%%%%%%%%%%%%%%%%%
\subsection{Violating the reasoning: Naturalness}

The concept of Naturalness was introduced by 't Hooft~\cite{tHooft:1979rat} in an attempt to distinguish acceptably and unacceptably large hierarchies within theories. In 't Hooft's original formulation a parameter $\eta$ is Natural if an enhanced symmetry arises as $\eta\to 0$. The reason for this is that quantum corrections respect symmetries, and so any nonzero contribution to $\eta$ must be a symmetry breaking contribution. The practical implication is that one sees evidence for naturally small parameters if the renormalization group equation is homogeneous in that parameter. That is, $\eta$ cannot be induced if $\eta=0$.

But Naturalness has gone far beyond 't Hooft's original formulation~\cite{Giudice:2008bi}. 't Hooft's original form is now called ``technical Naturalness" to separate it from what one might call ``finetuned Naturalness." Finetuned naturalness is when no observable quantities or parameter relations are highly finetuned with respect to the theory inputs. This is a somewhat fuzzy notion, whose power and utility can be tightened up by carefully defining an {\it a priori} conventional method of evaluation~\cite{Wells:2018yyb,Wells:2021zdp}.

Within our paradox formulation of the {\it Extrinsic Hierarchy Problem}, both technical and finetuned Naturalness are assumed as valid reasoning. This may sound controversial to some, since Naturalness has taken a beating in recent years as just a philosophical concept not rooted in science. However, the reasoning itself should not be viewed as controversial at all. Rather, it is the aleatory premise that is most difficult to assess. If we have parameters selected on a distribution, statistical tests, such as through finetuning determinations, can be applied warrantedly and productively.

In other words, it should be recognized that Naturalness has no utility outside of the aleatory premise. Without the aleatory premise no connection, not even a tenuous connection, can be made between finetuning and probability, rendering any finetuning assessments useless. Since we do assume the aleatory premise we accept Naturalness as a valid reasoning tool as long as we apply it to parameters selected from a distribution. 

This begs the question of whether we can say anything definitive using Naturalness reasoning if we do not know precisely the distribution of the underlying parameters, even if we admit that they are aleatory. The answer is yes in two senses. 

First, higher finetuning selects smaller regions of parameter space. Therefore, the higher the finetuning the lower the probability that those parameters can be selected. It is monotonic in that sense. Second, finetuning determinations are rather insensitive to the lower bounds and upper bounds on parameters considered in the analysis~\cite{Wells:2018yyb}. In the limit of flat distributions it is entirely insensitive, in contrast to other methods of reducing parameter space in an attempt to quantify decreasing probability.

There are many things in science where we do not know the precise probability densities, but if we assume that they are not wildly varying over some region of parameter space one can show relative probabilities rapidly diminishing as one goes to more and more restrictive parameter space. It is analogous to a flat tangent manifold being a correct description of space in a region of space much smaller than any characteristic variation.

We must recognize, however, that nothing in quantum field theory breaks if there exists an extreme finetuning. It is possible, but just not probable. Extreme finetunings are directly analogous to extreme coincidences (coincidences of equivalence between parameters or functions of parameters). In the past we did not settle for accepting the coincidence of the inertial mass being equal to the gravitational mass, or the lepton-decay fermi constant being accidentally the same as the nucleon-decay fermi constant. Extreme coincidences require principled explanations, and vice versa: typically only principled explanations can yield extreme coincidences (e.g., all the electromagnetic couplings being the same). For this reason, we view it likely that any new theories produced to resolve the {\it Extrinsic Hierarchy Problem} will adhere to exigencies of Naturalness reasoning.

%%%%%%%%%%%%%%%%%%%%%%%%%%%%%%%%%
\subsection{Interim Solutions}

There are a few theories that have been developed purportedly to solve the Hierarchy problem, either in full or in part, that arguably are immaterial to resolving the paradox as presented above. What these theories have in common is they construct plasters to cover up the blemish of nothing yet seen at the LHC, but they do not  cure  the theory's disease of being unstable to the highest scales of nature. Note, when we say ``theory" here, we mean whatever theory is before us: either the SM or a beyond-the-SM theory with its collection of fields and interactions that we want to know whether it is stable or not to generic expectations of additional states and interactions as we go up in the UV.

One example of such a direction are ``Little Higgs" theories. Little Higgs theories attempt to render the Higgs boson a pseudo-Nambu-Goldstone boson (pNGB). The benefit is that pNGB's have a shift symmetry $\phi\to\phi+a$ which keeps them massless to leading order. It also disables quadratic corrections to the Higgs boson mass since such corrections would create mass radiatively at one loop and would be accompanied by the perceived dangerous $\Lambda^2$ cutoff dependence of self-energy corrections.

One issue with Little Higgs theories is that they are trying to solve the old quadratic divergence $\Lambda^2$ problem within the SM theory --- the {\it Intrinsic Hierarchy Problem}. It partially does solve this problem, up to a scale that is a loop level or two loop levels above the weak scale. As discussed above, this may be a fake problem of no physical concern. 

The excellent review article of Little Higgs theories makes very clear that it is concerned entirely with canceling SM-induced quadratic divergences (i.e., the {\it Inrinsic Hierarchy Problem}):
\begin{quote}
{\it ``To understand the requirements on this new physics better we must look at the source of the Higgs mass instability. The three most dangerous radiative corrections to the Higgs mass in the Standard Model come from one-loop diagrams with top quarks, $SU(2)$ gauge bosons, and the Higgs itself running in the loop."}~\cite{Schmaltz:2005ky}
\end{quote}
Indeed, the entire purpose of the theory is to build an edifice that eliminates at the one-loop level any quadratic cutoff dependences of the Higgs mass due to SM particle loops.
In contrast, some view the radiative corrections of the Higgs mass from SM states to be of noconcern at all, and thus the Little Higgs theory sets out to solve a nonexistent problem.

Let us now ask whether the Little Higgs theory solves the {\it Extrinsic Hierarchy Problem}. This problem is formulated by applying all the premises and reasoning stated above to any particular Little Higgs theory and asking whether the theory remains Natural. One can immediately determine that the Little Higgs theories are a delicate balance of states and interactions, and that by adding a collection of other generic states that balance is disrupted.

Further emphasizing the point above, it is essential in the Little Higgs theories that the low-scale scalar states that condense and give rise to the Higgs boson as a pNGB do not couple to any other fundamental scalars at higher masses that could destabilize the entire setup. Scalars of high mass or high vacuum expectation values coupled to any of these Little Higgs states leads to  failure in its goal of solving the Hierarchy Problem. As soon as such states are admitted both the {\it Intrinsic} and {\it Extrinsic Hierarchy Problems} are immediately seen to be unsolved. In that sense, the theory is highly restrictive and nongeneric on the multitude of states it allows or its allowed couplings. Thus, it does not appear to solve the {\it Extrinsic Hierarchy Problem}. Indeed, we would not expect it to since it was never intended to solve that particular issue.

Little Higgs theories are clever and interesting, and they do partially solve a problem (the {\it Intrinsic Hierarchy Problem}), sometimes called the ``little hierarchy problem." Twin Higgs theories have similar motivations and were similarly never a full solution to the Hierarchy problem. Interim solution theories, such as Little Higgs and Twin Higgs, are invariably  highly vulnerable to the {\it Extrinsic Hierarchy Problem}. The theories push off the {\it central} fundamental problem to a higher scale, with the key aim of leaving human observers experimentally devoid of further clues for another decade or two of energy. Researchers of Little Higgs theories, and Twin Higgs theories, understood this well and invoked more absolute means to stabilize at higher scales. For example, some introduced supersymmetry~\cite{Katz:2003sn,McCullough:2025zeg} and others extra dimensions~\cite{Contino:2003ve,Thaler:2005en} to give the ultimate stabilizing mechanism, and we are back to the more fundamental solutions discussed above.

%%%%%%%%%%%%%%%%%%%%%%%%%%%%%%%%%%%%%%%%%%%%%%%%%%%%%%%
%%%%%%%%%%%%%%%%%%%%%%%%%%%%%%%%%%%%%%%%%%%%%%%%%%%%%%%
%%%%%%%%%%%%%%%%%%%%%%%%%%%%%%%%%%%%%%%%%%%%%%%%%%%%%%%
\msection{Conclusions}

In conclusion, we have discussed the Hierarchy Problem in elementary particle physics, and we found it fruitful to separate it into two related but distinct problems: the {\it Intrinsic Hierarchy Problem}, and the {\it Extrinsic Hierarchy Problem}.

The {\it Intrinsic Hierarchy Problem} takes seriously a physical interpretation of the cutoff scale in Wilsonian renormalization group. This immediately creates a significant problem for any interacting scalar field theory whose mass is hierarchical smaller than the cutoff scale. It constitutes an {\it intrinsic} problem for the theory. However, we have argued that this may be a faux problem, connected to an arbitrary cutoff choice in how one regulates the infinities of a quantum field theory. Nevertheless, it remains a legitimate concern and current area of inquiry on the interpretation and implications of Wilsonian effective field theories.

The {\it Extrinsic Hierarchy Problem} takes seriously the generic expectation that Nature has many more particles at every accessible energy scale, and these particles are generically expected to couple to all other states any way they are allowed. This creates a destabilizing impact on the Higgs boson, which through its gauge invariant and Lorentz invariant dimension-two operator $|H|^2$ can coupling to any number of other states through quantum loops or even directly $|H|^2|\Phi_i|^2$.

We showed how the {\it Extrinsic Hierarchy Problem} is well suited to be phrased as a paradox. The premises we identified for the paradox are 1) conventional Ur-Theory, 2) aleatory parameters, and 3) multitude of states. We assessed many ideas that purport to solve the Hierarchy Problem and were able to catalog different theories according to which premise they attacked. We also identified theories that have been constructed to solve the Hierarchy Problem, but in fact have limited effectiveness for the {\it Intrinsic Hierarchy Problem} and no effectiveness for addressing the {\it Extrinsic Hierarchy Problem}.

The lack of evidence of new physics at the LHC at mass scales in the neighborhood of the Higgs boson is concerning for advocates of theories that directly solve both Hierarchy Problems.  We argue that although it may be too early to give up on those classic theories, we may soon need to abandon the Wilsonian dogma of UV-IR independence altogether if we are to resolve the Hierarchy Problem, either the Intrinsic or Extrinsic versions.

%%%%%%%%%%%%%%%%%%%%%%%%%%%%%%%%%%%%%%%%%%%%%%%%%%%%%%%
%%%%%%%%%%%%%%%%%%%%%%%%%%%%%%%%%%%%%%%%%%%%%%%%%%%%%%%
%%%%%%%%%%%%%%%%%%%%%%%%%%%%%%%%%%%%%%%%%%%%%%%%%%%%%%%
\bigskip
\noindent
{\it Acknowledgements:} I thank R.\ Akhoury and S.\ Martin for helpful discussions on these topics. I also wish to thank the organizers and participants of the ``New Approaches to Naturalness" conference (Lyon, France, May 2025), where this work was first presented~\cite{Wells:2025-Lyon}, for stimulating discussions. Support from LITP is gratefully acknowledged.

%%%%%%%%%%%%%%%%%%%%%%%%%%%%%%%%%%%%%%%%%%%%%%%%%%%%%%%
%%%%%%%%%%%%%%%%%%%%%%%%%%%%%%%%%%%%%%%%%%%%%%%%%%%%%%%

\end{document}